\documentstyle[preprint,aps]{revtex}
\begin{document}

\preprint{\vbox{\baselineskip=15pt
\hbox{NSF-ITP-93-85} \hbox{gr-qc/9307010}}}

\title{Late-time behavior of stellar collapse and explosions:
\\II. Nonlinear
evolution}

\author{Carsten Gundlach, Richard H. Price and Jorge Pullin}
\address{Department of Physics, University of Utah, Salt Lake City, UT
84112-1195}

\date{9 July, 1993}

\maketitle

\begin{abstract}
We compare the predictions of linearized theory for the radiation
produced in the collapse of a spherically symmetric scalar field with
a full numerical integration of the Einstein equations. We find
power-law tails and quasinormal ringing remarkably similar to
predictions of linearized theory even in cases where nonlinearities
are crucial. We also show that power-law tails develop even when the
collapsing scalar field fails to produce a black hole.

\end{abstract}

\pacs{04.30.+x, 04.40.+c}


\section{Introduction}

Linearized perturbation theory has been the main analytical -- and
until comparatively recently, numerical -- tool for analyzing
nonspherical gravitational collapse.  The complexity of the problem has
usually made this approach necessary, and it has been assumed until
recently that the approach was sufficient.  Recently, however, G\'omez
and Winicour \cite{GoWi} have focused attention on the extent to which
these results are even qualitatively representative of the late stages
of collapse.

In the picture given by linear perturbation theory of the late stages
of collapse, there are two features which are noteworthy. One is the
development of ``quasinormal (QN) oscillations,'' damped oscillations
at complex frequencies characteristic of the mass of the black hole
background.  The second feature is the decay in time $t$ of
perturbations as $1/t^n$, the ``power-law tails.'' There are good
reasons to examine more carefully whether these features also appear
in the fully nonlinear case. The arguments for the QN oscillations and
for the tails are somewhat different, and should be considered
separately.

According to linearized theory the QN frequencies are fixed complex
numbers multiplied by the inverse of the mass of the black hole
background.  (We use here and throughout units in which $c=G=1$.) 
It seems reasonable that the phenomenon of QN ringing will be a
feature of nonlinear collapse. One argument is that some numerical
investigations of solutions of the fully nonlinear equations have
shown QN ringing to be common \cite{AbBeHoSeSm}. Secondly, the idea of
QN ringing seems ``robust.'' It is a natural frequency associated with
a radiative boundary condition, and can occur in many different
radiative systems.  If QN ringing is found, to what black hole mass
does it correspond? The QN oscillations themselves carry energy and
may change the meaning of the mass. A reasonable guess, at least, can
be made that the QN frequency evolves somewhat during the collapse.

The situation for the power-law tails is quite different. These tails
are not familiar or common phenomena. The explanation of their
existence can be given in two very different ways: (i) They can be
viewed as the result of the scattering of gravitational waves off the
``effective curvature potential'' of the black hole
spacetime \cite{Pr}, or (ii) they can be associated
with the branch cut in the Green function for the wave propagation
problem \cite{Le}. Both arguments leave open the possibility that
the tails are idiosyncrasies of the linear approximation.

If QN oscillations or tails are missing from a fully nonlinear
collapse, or if there is any significant new qualitative feature, the
result might be to undermine confidence in the picture of collapse
given to us by the analysis of linear perturbations of black hole
backgrounds.  G\'omez and Winicour \cite{GoWi} have addressed this
question with numerical studies of the collapse of a spherically
symmetric scalar field due to its own gravitational pull. Since the
spherically symmetric problem involves only a 1+1 hyperbolic system,
it is enormously easier to solve numerically than the problem of
nonspherical collapse.

What is more, the problem is a wonderful testing ground for comparing nonlinear
results and the predictions of linearized theory. In linearized
perturbation theory the evolution of a scalar field is governed by
essentially the same mathematics that governs the dynamics of
nonspherical perturbations. In particular, perturbation theory makes
very specific predictions about QN ringing and tails for perturbative
spherically symmetric scalar fields.  It is therefore of great
interest that in their initial numerical studies of scalar field
collapse, G\'omez and Winicour have seen neither QN ringing nor tails.

In this paper we will study the fully nonlinear evolution of a scalar
field minimally coupled to general relativity. We consider first the
evolution of a spherically collapsing scalar field; in addition, we
consider nonspherical perturbations of this spacetime. We establish
that the QN frequencies and the power-law tails of the numerical
simulations are in remarkable agreement with the predictions of
linearized theory when a black hole develops. If a black hole does not
develop, and all energy eventually radiates away to infinity, we find
that power-law tails still form. The existence of tails, but not QN
oscillations, when holes do not form agrees with the analysis
presented in the companion paper, hereafter referred to as Paper I.  

The organization of this paper is as follows. In section II we
describe the coordinate system and the version of the field equations
we use. In section III we describe our discretization and discuss the
numerical error.  In section IV we study the collapse of a scalar
field for various initial configurations. In section V we study the
evolution of multipole moments of test fields on the  collapsing
background. In all cases comparisons with the linearized results are
made. We end in section VI with a brief summary and with conclusions.


\section{Field equations and algorithm}

We study a spherically symmetric scalar field $\phi$ satisfying the
minimally coupled equation
\begin{equation}
\label{0} \phi^{;\mu }_{\ ;\mu }=0.
\end{equation}
To describe the spherically symmetric spacetime we use the line element
\cite{Chris,GoPi},
\begin{equation}
\label{1}
ds^2\equiv-g(u,r)\bar g(u,r)\,du^2-2g(u,r)\,du\,dr-r^2\,d\Omega^2,
\end{equation}
in which $u$ is a retarded time null coordinate. Regularity at the
center requires that $g=\bar g$ at $r=0$. The remaining coordinate
freedom is fixed by the choice that $u$ be the proper time on the
$r=0$ central world line, or $g=\bar g=1$ at $r=0$.  We introduce the
auxiliary field $\psi$ by
\begin{equation}
\label{2}
\phi\equiv{1\over r}\int_0^r \psi\,dr.
\end{equation}
In terms of this variable, the wave equation (\ref{0}) for $\phi$ takes the form
\begin{equation}
\label{3}
\left({\partial\over \partial u}-{\bar g\over2}{\partial\over \partial
r}\right)\psi\equiv D\psi={1\over2r}(g-\bar g)(\psi-\phi).
\end{equation}
The metric coefficients $g$ and $\bar g$ are determined from 
$\phi$ by
\begin{equation}
\label{4}
g=\exp\left[4\pi\int_0^r 
\left({\partial\phi\over\partial r}\right)^2 r\,dr\right],
\end{equation}
\begin{equation}
\label{5}
\bar g={1\over r}\int_0^r g\,dr.
\end{equation} 
Because of the spherical symmetry there is no independent
gravitational degree of freedom.
The operator $D$ in (\ref{3}) differentiates along ingoing null
characteristics, and information is propagated along outgoing null
characteristics (the $u=$ constant lines) by (\ref{2}). The
integration scheme of equations (\ref{2}) and (\ref{3}) can therefore
be 
related to a  fully characteristic coordinate system
($u=$``retarded time'', $v= $``advanced time,'') in which the metric takes the form
\begin{equation}
\label{6}
ds^2=-f(u,v)\,du\,dv+r^2(u,v)\,d\Omega^2 ,
\end{equation}
and in which $D\propto\partial/\partial v|_{u}$. Here the coordinate
$v$ is fixed only up to arbitrary transformations
$v\rightarrow\tilde{v}(v)$. Since the coordinate is only a label on
the ingoing null geodesics it can be assigned values $v=1,2,3...$ on
our numerical grid.

Our algorithm works on a characteristic grid made up of lines of
constant $u$ and $v$, with the radial ``coordinate'' $r$ treated as
a metric function $r(u,v)$ which is evolved as
\begin{equation}
\label{8}
D\,r=-\bar{g}/2\ .
\end{equation}
As  initial data for the algorithm we take $\psi(v)$ and $r(v)$ on
an initial outgoing null cone $u=u_0$.  These are equivalent to the
choice of null data $\phi(r)$ and the (arbitrary) numerical choice of the
initial position in $r$ of each ingoing null lines of the grid (i.e.,
each line of constant $v$).  The algorithm, which closely resembles
that of Goldwirth and Piran \cite{GoPi}, proceeds as follows: We start
by using (\ref{2}) to obtain $\phi$ and by using (\ref{4}) and
(\ref{5}) to obtain $g$ and $\bar g$ as functions of $v$ along
$u=u_0$. (The integrations over $r$ are discretized as summations over
$v$.)  We next choose a value of the ``time step'' $\Delta u$, and we
use (\ref{3}) to obtain $\psi$, and (\ref{8}) to obtain $r$, at grid
points along $u=u_0+\Delta u$. The process is then repeated starting
with the new outgoing characteristic $u=u_0+\Delta u$.  On each such
cycle of integrations the step size $\Delta u$ is chosen so that
\begin{equation}
\label{Courant}
|r(v,u+\Delta
u)-r(v,u)|\leq |r(v+1,u)-r(v,u)|
\end{equation}
for all $v=1,2,3,...$ This has been
found to be a useful rule of thumb.  Since our integration scheme
proceeds along characteristics, there can be no violation of
causality, and hence there is no formal ``Courant condition'' to
satisfy.

Figure 1 shows our $uv$ grid embedded in the well-known conformal
diagram of spherical collapse. Null lines $u=const.$ and $v=const.$
are at 45 degrees. Infinity has been brought to a finite distance and
each point, apart from the line $r=0$, corresponds to a 2-sphere of
surface $4\pi r^2$. This diagram shows where our null data are set and
how far they are evolved. The upper left and right sides of our
coordinate patch, though at finite distances, can be taken as
approximations of the horizon ${\cal H}_+$ and future null infinity
$scri_+$. Lines of constant $r$ go from past to future timelike
infinity. Those with $r<2M_f$, where $M_f$ is the final mass of the
black hole, cross the event horizon, while $r=2M_{f}$ approaches it
asymptotically. This gives one numerical method for finding the final
black hole mass. Another is to take the limit of the Bondi mass
$M_B(u)$ at late retarded times $u\to\infty$, where
\begin{equation}
M_B(u)\equiv\lim_{r\to\infty}\,\left(1-{\bar g(u,r)\over g(u,r)}\right)
r^2\,.
\end{equation}

Yet another spacetime coordinate will be useful in comparing our
results to analytic predictions, although it plays no active role in
our algorithm.  This additional coordinate is defined as the proper
time along an $r=const.$ trajectory, or
\begin{equation}
t=t(u,r)\equiv\int_0^u \sqrt{g(u',r)\bar g(u',r)}\,du'.
\end{equation}
We shall also use Bondi time, $t_B\equiv t(u,\infty)$,
which is the retarded time coordinate that agrees with
time at infinity for constant $r$. In an
asymptotically flat spacetime the large $r$ limit of $t(r)$ is given
by $dt_B=\lim_{r\to\infty}\bar g(r,u)\, du$.

Our code has limited integration time precisely because it is
characteristic.  By definition $u$ is the proper time of an observer
at $r=0$. The event horizon starts spreading out from $r=0$ at a
finite value of $u$, say $u_h$.  In a collapse $\bar g(\infty,u)$
becomes infinite as $u\to u_h$.  In our algorithm this means that the
stepsize $\Delta u$ decreases rapidly, while $\Delta t_B$ remains
about constant. The numerical approach to the horizon is stopped
eventually by an overflow of $\bar g$ or underflow of $\Delta u$. The
situation is best illustrated in Fig.~2.  The $u=const.$ lines of our
grid are squeezed together against the horizon.  The problem of
over/underflow, and of numerical instability (which usually develops
earlier) is shared by all codes which avoid singularities by staying
outside apparent horizons. A possible cure would be a slicing which
does cross the horizon, combined with simply discarding parts of each
slice inside the horizon. If the surface beyond which time slices are
discarded eats into the remaining part of the grid with the speed of
light or faster, no boundary condition is required on it \cite{SuSei}.
Such a procedure is not applicable to a characteristic grid, where the
``time'' slice is already an outgoing null cone and cannot be
intersected by one.


\section{Discretization and error analysis}

As we have just seen, our grid is highly nonuniform in $u$ if a
horizon forms.  In that case it becomes also highly nonuniform in $r$.
Ingoing null geodesics pile up at $r=2M_f$, even if $v$ is chosen
uniform in $r$ on the original slice.  Truncation error is reduced, of
course, if the grid spacing is made finer.  We have tried to produce a
code that is second-order accurate under a uniform grid rescaling.  We
denote here the relative size of the grid spacing by $h$; a reduction
by a factor 2 of $h$ means that the spacing in $r$ is reduced by 2
and, due to the Courant-like condition (\ref{Courant}), the spacing
$\Delta u$ is also reduced by 2.

 On
the top level we treat each of equations (\ref{3}) and (\ref{8}) as an
ordinary differential equation in $u$ at each fixed $v$. These are
solved by the second-order Runge-Kutta method.  The calculation of the
``coefficients" $\phi$, $r$, $g$ and $\bar g$ in these ``ODEs" is
nontrivial however, and couples the equations at different values of
$v$. They are given by the definite integrals (\ref{2}), (\ref{4}) and
(\ref{5}), which must be evaluated in that order. The integrals are
discretized by the trapezoidal rule, which is again formally
second-order accurate.  

In principle our code should be second-order. That is, for the finite
difference approach we use, the error at a given spacetime point
should vary as $h^{2}$.  We did not find this second-order convergence
one would expect naively from the code, but did find convergence
better than first order. One reason for the failure of a simple error
analysis is the ambiguity about the measure of error. As the dynamical
range of $\phi$ and the other fields is very large in a collapse
spacetime, there is little point in taking an $l_2$, $l_1$ or
$l_\infty$ norm of the error in the fields over a set of spacetime
points covering all of the integration region.  The regions where the
fields are large would  dominate the integrated error. What
is relevant to our confidence in the results is the {\em relative}
error, in particular in the regions where we are looking for tails and
where $\phi$ is small.

Because of this we have looked at small spacetime regions, taking the
$l_2$ norm over only a few neighboring points. For all these regions
we found power-law dependence of errors to apply over a change in
grid size by a factor of 16.  Roughly speaking, error scaled as
$h^{N}$, with $N=1$ for small $r$ or small $t_B$, and
increasing up to 1.6 for both $r$ and $t_B$
large. 
By small $r$ we mean the very narrow shell in $r$, just outside
$r=2M_f$, in which $\phi$ remains large at all times and where its
gradient increases without bound. By small $t_B$ we mean the
region where the bulk of outgoing radiation passes.

Figures 3 and 4 display the error in of $\phi(r,t_B)$, for sections of
the lines $r=10$ and $t_B=50$. In these figures the initial data is a
Gaussian $\phi(r,t_B=0)$ with center $1.0$, width $0.1$ and amplitude
$0.06$.  (See the next section for a general discussion of initial
data.)  As there are no useful analytical solutions available, we had
to use a numerical solution as the reference solution. We chose a
solution generated with an initial grid spacing of $\Delta r=1/320$,
and compared to it grid spacings of $1/20$, $1/40$, $1/60$, $1/80$,
$1/120$, $1/160$ and $1/200$. (In all of these, however, the grid
spacing was smoothly reduced by another factor of $16$ for grid values
of $0<r<1$, as we shall explain below.)  The values of $\phi$
for the lower-resolution runs were interpolated linearly to the values
of $r$ and $t_B$ of the grid points of the highest-resolution run. The
difference of $\phi$ between a run and the reference run was squared,
summed and the square root taken.

Figures 5 and 6 establish visually that the code converges, by showing
sections of $\phi(r=10,t_B)$ for different grid spacing. Figure 5 shows
the region dominated by QN  ringing, and Fig.~6 a region
dominated by the power-law tail.

It is perhaps surprising that the error varies as $h^{N}$, even if $N$
varies over spacetime and does not have the value 2 that one might
suppose. The fact, however, that this power is closer to 1 both near
the origin and in regions of large $\phi$, and closer to 2 elsewhere,
seems to be an indication that the error is largely due to our
handling of the $r=0$ boundary conditions.

We were alerted by reference \cite{ChGoPi} to the possibility that the
main source of error may be the boundary conditions $\bar g=g$ and
$\phi=\psi$ at $r=0$. It can be seen in Fig.~1 that one after another,
the $v=const.$ grid lines cross $r=0$ into unphysical negative values
of $r$, at which point they are dropped from the algorithm. We have
numerically implemented the $r=0$ boundary condition by approximating the
true value of $\psi(u,r=0)$ by linear interpolation. Formally this is
only first-order accurate. The situation is more involved, however:
the right-hand side of (\ref{3}) contains an explicit factor of $1/r$,
which in the exact solution is cancelled by the very boundary
condition we are trying to impose, and which analytically leads to
$g-\bar g=O(r)$ and $\psi-\phi=O(r)$. The risk of large numerical
error, or even numerical instability, is clear. Empirically we found
that our linear approximation algorithm is stable, but gives rise to
large errors at small $u$, which actually decrease with increasing
$u$. We believe the reason is that at small $u$ the scalar field
$\phi$ is still large at $r=0$, thus introducing a large error into
the boundary condition. At large $u$, the field strength resides
mostly at large $r$, and in this regime the standard discretization
error is dominant. We addressed this problem, not by using
higher-order interpolation, but by brute force. We made the grid much
denser in $r$, typically by a factor of 16, up to and slightly beyond
the radial scale of the initial data. This meant that the grid size
was much smaller until the bulk of the energy in the scalar field had
reached $r=0$ and then had propagated out to
large $r$.  In this way we achieved a reasonably small relative error
everywhere, small enough to give us confidence in our results.

In summary, our code is not second-order accurate, but better than 
first-order accurate. We achieved high enough accuracy to have a small
relative error everywhere, but even at a much lower resolution
the physical features we had set out to verify,
QN ringing and power-law tails, are unambiguously present.

\section{Spherical collapse}

In this section we present our results for the purely spherical
collapse of a self-gravitating minimally coupled scalar field. Our
expectations are based on linearized perturbations as analyzed in the
Paper I.  Because the scalar field at late times has small amplitude,
it seems plausible that the late-time fields can be viewed, roughly,
as perturbations, and that the analysis of Paper I applies, at least
approximately. Our expectations then include the presence of QN
oscillations at late times after a black hole has formed, and
power-law tails at late times, both when a black hole forms and when
it does not. The real and imaginary part of the QN frequencies -- the
least damped mode is the only one visible in practice -- are fixed
numbers divided by the black hole mass. The powers of the linear
theory tail are fixed integers. When there is a static scalar field
present on the initial null slice, the amplitude of the power-law
tail, in the linearized case, is also determined. The analysis in
Paper I suggests that these predictions for a scalar test field should
hold also, approximatively, if the spherically symmetric scalar field
evolves under the influence of its own gravitational pull rather than
on a fixed Schwarzschild background.

We examined two one-parameter families of initial data. In the first
family the field is a Gaussian in $r$; we consider the Gaussian to
represent a typical collapsing ``shell of matter.''  There is a scale
invariance in the problem which can be used to set the center of the
Gaussian at $r=1$; the remaining physical parameters are the Gaussian
width and amplitude.  A black hole will form from these initial data
if the amplitude is sufficiently large, or if the Gaussian width is
sufficiently small.

As a second family of data, two different forms of $\phi(u=0,r)$ are
joined. For $r$ less than some joining radius $r_{\rm J}$, the
solution is constant at $\phi=\phi_{0}$, and for $r>r_{\rm J}$, the
initial data is taken to be that static solution of (\ref{0}) which is
well behaved as $r\rightarrow\infty$ (i.e., the solution which falls
of as $r^{-1}$).  This solution is called ``static-static'' \cite{GoWi}
because it gives rise to a static spacetime in the domain of
dependence of either the inner ($r<r_{\rm J}$) or outer ($r>r_{\rm
J}$) initial data.
Without loss of generality we can set the joining point at $r_{J}=1$.
The only free parameter is then $A$, the field amplitude at that point.
Again this one-parameter family should contain spacetimes with and
without an event horizon.

Static-static data are the boundary between generic ``shell" initial
data and data which are not asymptotically flat. We found that when we
replaced the exact static solution (given analytically as an implicit
function in \cite{Buch,GoWi}) by $\phi=const./r$) the resulting
solution is qualitatively different: it corresponds to an infall of
matter at arbitrarily late times.

A first check of our code is the amplitude of static-static data at
which a black hole first forms. We find $0.28<\phi_0<0.29$, which
agrees with the critical value $\phi_0=0.286$ given in \cite{GoWi}.
For the Gaussian data we find $0.033<A<0.034$, where $A$ is the
height of the Gaussian. These limits are stable under reduction of the
grid size by a factor of 16, from 1/20 down to 1/320, as described in
the previous section.

A crucial feature of both Gaussian and static static data is that the
amplitude (the value of $\phi_{0}$ or the Gaussian height $A$) can be
chosen either ``subcritical'' (no hole formation) or ``supercritical''
(a hole forms). By varying the amplitudes we
found three different regimes for the resulting spacetimes.  a) For
the amplitude much greater than its critical value we found QN
oscillations and tails qualitatively and, for the most part
quantitatively, as predicted by linearized analysis.  b) For the
amplitude near its critical value we found the same power-law tails,
but we found QN ringing to be qualitatively different in the Gaussian
and static-static cases.  c) For the amplitude smaller than its
critical value (even marginally) we found
the same power-law tails as for collapse, but no QN ringing.

We now discuss in more detail the results for power-law tails and for
QN ringing, starting with the former.  The analysis in Paper I shows
that for linear spherical perturbations tails should fall off in time
(time for a distant observer, or Bondi time) as $t_{B}^{-3}$ in the
case of generic data, and as $t_{B}^{-2}$ for data corresponding to an
initially static monopole moment. Figure 7 shows log-log plots of
$\phi$ at a constant values of the radius (here $r=10$) as a function
of Bondi time for different values of the amplitude of the initial
Gaussian data.  At late times $\phi$ clearly decays as a power of
Bondi time, $t_B^{-n}$. The measured exponents are in reasonably good
agreement with the prediction of linearized theory: $n=2$ for static
static data, and $n=3$ for Gaussian data. These power-law tails
develop, and have the same exponents, whether or not a hole forms, and
the amplitude of the tails is a smooth function of the initial
amplitude.  For the development of a tail, hole formation is
irrelevant. This independence of hole formation is explained by the
argument in Paper I that tail development is due to backscatter at
large $r$, and does not depend on the small $r$ nature of the
spacetime.

In the case of static-static data, the linear perturbation analysis of
Paper I predicts not only the exponent of the tail, but the amplitude.
If our static static data were evolving on a fixed Schwarzschild
background of mass $M$, the Paper I prediction would be that at late
times and large radii (specifically $t\gg r\gg M$) the tail would have
the form $\phi=-4M\phi_0 /t^2$. It is unclear just how to apply this
in general to the time dependent spacetimes of our present numerical
investigations. Since the mass of the spacetime is evolving, and might
fall to zero, what value of ``$M$'' should be used in the prediction
for the amplitude of the tail? The picture of scattering underlying
the calculations of Paper I would suggest that the correct $M$ is some
average, over retarded time, of the mass. The question of the
appropriate mass is not crucial in one subset of cases: collapse
models in which the initial data leads to the formation of a hole of
mass not very different from the initial Bondi mass.  In this case one
might expect the linear fixed background prediction to apply. Our
results, however, do not show this. In all cases, whether subcritical
or supercritical, the amplitude of the tail is an order of magnitude
less than the prediction of linearized theory (using the initial Bondi
mass for $M$ in the above expression).  The reason for this is
not yet clear.

The results for QN ringing differ in an important way from those for
tails.  QN ringing is a late time oscillation associated with the
``effective potential'' governing the dynamics of perturbation fields
outside a hole. Unlike the power-law tails, QN ringing depends on the
small $r$ nature of the spacetime. If a hole (or relativistically
compact object) does not form, the generation of QN oscillations is
not expected. This is verified by the results; we have found no QN
ringing for solutions without black holes. The transition from
solutions with QN ringing to solutions without is, however,
qualitatively different for Gaussian and static-static data.

For Gaussian data, this transition is abrupt; QN ringing is clearly
present in a marginally collapsing spacetime, and clearly absent in a
marginally noncollapsing spacetime. This is strikingly illustrated in
Fig.~8, showing the solutions evolved from Gaussian data with
amplitudes $0.033$ and $0.034$.  We plot $\phi$ as a function of Bondi time at
constant radius. Initially the two solutions are close, as one would
expect from the closeness of their initial data.  Later, the collapsing
solution is delayed with respect to the noncollapsing one, and shows
QN ringing. But after QN ringing has died away, and the power law tail
dominates both solutions, they are again close.  This result is
compatible with the linearized analysis given in Paper I.  Power-law
tails are the result of backscatter, at large $r$, of the initial
outgoing burst of waves. The two near-critical curves in Fig.~8 have
almost identical outgoing bursts (since they correspond to very nearly
the same initial data) so, at late times, the tails should be the
same. The QN ringing, on the other hand, depends on how the solution
develops at small $r$ and hence is expected to be very different for
the subcritical and supercritical cases.

For static-static data, the QN ringing fades away gradually before the
critical amplitude is reached from above. In consequence,
$\phi(r=10,t_B)$ is nearly the same function everywhere for marginally
collapsing and noncollapsing data; see Figs. 9 and 10.  It may be that
the QN ringing is masked, in the marginally collapsing case, by the
 non-QN fields, but the explanation at present is not
clear.

It is difficult to measure the QN frequencies for the spherical scalar
field, as only one minimum and one maximum is visible. For an initial
Gaussian of amplitude $0.06$ for example, one measures a half-period
of about $4.0$. In this case the initial Bondi mass is $0.1699$ and
the final Bondi mass still $0.1696$, so that the mass is practically
constant. The half-period predicted for spherical scalar
perturbations\cite{LeavThs} is $28.44 M$, so on the basis of linear
perturbation theory we would expect a half period between $4.832$ and
$4.824$, which agrees with what is seen in the numerical results
within the rather unsatisfactory precision available.


\section{Perturbations of spherical collapse}

To understand better the results of the nonlinear evolution of the
spherically symmetric scalar field, we have investigated a closely
related problem. We consider a second minimally coupled scalar field
$\varphi$, not coupled to the $\phi$ field, and we treat this second field
perturbatively. That is, we ignore the contribution of $\varphi$ to the
stress-energy, and hence to the spacetime geometry. We study, then,
the evolution of the perturbative $\varphi$ field on the backgrounds
generated by collapsing $\phi$ fields.  Since these backgrounds are
spherically symmetric, the perturbations can be decomposed into
spherical harmonics which decouple. The equations of motion for a
perturbation of fixed angular dependence
$\varphi=\varphi_l^m(u,r)Y_l^m(\theta,\phi)$ are simply
\begin{equation}\label{Dpsi}
D\psi_l^m={1\over2r}(g-\bar g)(\psi_l^m-\varphi_l^m)
+{1\over2r}l(l+1)g\varphi_l^m,
\end{equation}\begin{equation}
\varphi_l^m\equiv{1\over r}\int_0^r \psi_l^m\,dr,
\end{equation}
where $g$ and $\bar g$ are still determined by the 
background solutions for $\phi$, through (\ref{4}) and (\ref{5}). The last term in (\ref{Dpsi}) turns out to drive instabilities near
$r=0$ for $l\ne 0$. The simple expedient of reducing the step size
$\Delta u$ to a value much smaller (we used 1/16) than that of
condition (\ref{Courant}) was found to remove the instability.  (We
are grateful to Jeffrey Winicour for suggesting this remedy.)

It should be understood that the equations governing $\varphi$ are not
quite the same as those that would describe perturbations of the
$\phi$ field itself. The stress-energy tensor is quadratic in $\phi$,
so that a perturbation $\delta\phi$ in $\phi$ would give rise to a
perturbation in the stress-energy tensor that is first order in
$\delta\phi$, and proportional to the ``background'' value of $\phi$.
It follows that there would be a first order perturbation in the
spacetime geometry. The equations governing $\delta\phi$ would
therefore {\em not} be the equations for a minimally coupled field on
the background spacetime given by the background $\phi$ solution. (The
situation is similar to that for electromagnetic perturbations of the
Reissner-Nordstr\"om spacetime. Since there is a background
electromagnetic field, the equations for electromagnetic perturbations
are not simply the Maxwell equations on the Reissner-Nordstr\"om
spacetime.) Despite this,  one should expect that there is
not a great deal of difference in the {\em late time} behavior of
$\varphi$ and of $\delta\phi$. Since the background $\phi$ field becomes
very small at late times, the stress-energy perturbations 
should be very small.

The motivation for studying the $\varphi$ field is that the scalar
collapse models go beyond the models of Paper I in two distinct ways.
First, the scalar collapses involve nonlinear field evolution, and
second, they produce time dependent spacetime geometries on which the
tails and the QN oscillations are developing. The introduction of a
perturbation field allows us to separate these two complications. The
dynamics of $\varphi$ is purely linear, but takes place on the time
dependent background spacetime. A study of the $\varphi$ field has an
additional attraction: For fixed backgrounds the exponents
of the power-law tails depend on the multipole index $l$. For generic
data the tails fall off as $t^{-(2l+3)}$; for initially static
multipoles they fall off as $t^{-(2l+2)}$. By computing different
multipoles of $\varphi$ we can check whether this $l$ dependence applies
also to time dependent backgrounds.  A further technical advantage is
that multipole modes of higher $l$ have QN frequencies which are
considerably less damped than the least damped $l=0$ mode.  This
allows us much more readily to ``observe'' the presence  of QN
oscillations and to measure their frequencies.

As an example we turn again to the solution with Gaussian data of
amplitude $0.06$, an example we considered in the previous section. For
the $\varphi$ test field we choose a Gaussian of width $0.1$, the same
width as for the background $\phi$ field, but we put the center of the
$\varphi$ Gaussian at $2.0$ while the center of the $\phi$ Gaussian,
which generates the background, is at $1.0$. This means that the
perturbation ``shell'' goes in later than the background shell, giving
the black hole time to form. For the $l=0$ $\varphi$ field we again (as
in section IV) find a half-period of about $4.0$, and again find it
difficult to determine the period with any accuracy, so that a
meaningful comparison cannot be made with the fixed background
prediction of $4.824$ (calculated from a final Bondi mass of
$0.1696$). For the $l=1$ $\varphi$ field, however, we could see 10
oscillations, and could measure a half-period of $1.811$, which is
within $0.7$ \% of the predicted value of $1.819$. For $l=2$ there are
20 oscillations, with the measured half-period of $1.105$, within
$0.2$ \% of the predicted value of $1.104$. QN ringing for these cases
is shown in Fig.~11.  

We also examined $l=2$ QN ringing on a background with considerable
mass loss during the time of QN excitation. To arrange this a Gaussian
``shell'' of $\phi$ and a Gaussian ``shell'' of test field $\varphi$
were both centered at $r=1.0$.
The $\phi$ field was chosen to have amplitude $0.033$. This  gives an
initial Bondi mass of $0.05822$ which decreases to a final Bondi mass
of $0.02035$, with most of the mass loss taking place during
$2<t_B<4$. For these masses, linearized theory predicts that  
the frequency of the least
damped $l=2$ mode should go up
from $8.31$ to $23.87$. This is in excellent agreement
with our numerical results, within the inevitable uncertainty in
measuring a varying frequency.

For power-law tails the most interesting backgrounds are those that
are the most time dependent, the noncollapsing backgrounds.  For these
backgrounds results are shown in Fig.~12 for tails from Gaussian data
with an amplitude of $0.020$ and in Fig.~13 for tails from
static-static data with an amplitude of $0.10$. As these figures
show, the late time behavior is rather accurately that of a power-law
tail; the numerically determined exponents are in rough agreement with
the predictions of linearized theory, and show the predicted increase
of slope with multipole index $l$.


\section{Discussion}

In Paper I we revisited the argument \cite{Pr} for the existence of
power-law tails of perturbations of Schwarzschild spacetime, in order
to point out that it predicts these tails not only at timelike
infinity, but also at null infinity and on the horizon.  Reformulating
the argument yielded an important spin-off prediction: Power-law tails
should form in general not only in perturbations of black holes. The
analysis suggested that the tails should form for massless
perturbations of any approximately spherically symmetric spacetime
whose metric is approximately Schwarzschild, at least on an outgoing
null cone of finite thickness (i.e., finite range of advanced time).

The central idea, as elaborated in Paper I, is quite simple. Radiation
going out along a thick null cone (the ``initial burst'' of radiation)
will be scattered back in by the spacetime curvature at arbitrarily
large radius.  The backscattered radiation then reaches a small radius
again at arbitrarily large time, attenuated by a power of Bondi time.
This backscattered radiation then evolves tails.  The leading effect
in this evolution is independent of spacetime curvature. The exponent
of the tails is therefore the same if at late times 
there is, at small $r$, a star, empty space, or a black hole.

In our numerical work described here we first set out to verify the
existence of the tails, as well as of QN ringing, in a collapse
situation. We chose the model of a spherically symmetric
self-gravitating massless minimally coupled scalar field, because
tails and QN oscillations seemed to be absent in the results of a
previous investigation\cite{GoWi}. The explanation of that absence
seems to be simply that one has to go to very late times to see the
late-time features. We did find tails and QN ringing precisely as
expected. Fortunately these features arise about an order of magnitude
in time before our code is stopped by a diverging redshift.

Extending the scope of previous work, we paid attention to the
late-time behavior of solutions which are subcritical, just below or
well below the margin of black hole formation, and we again found
power-law tails, but not QN ringing. For what can be considered
``generic" data (see Paper I), we found that QN ringing disappeared
abruptly at the margin of collapse, while the appearance of power-law
tails was continuous and unremarkable across  the transition from
subcritical to supercritical models. This last observation in itself is
a very strong corroboration of the simple picture of tail formation set
out above. Further evidence can be found in the dependence of the tail
amplitude on the amplitude of the initial data, as shown in Fig.~14.
The Bondi mass of the spacetime scales as the square of the amplitude
of the initial data and we find that the tail amplitude scales as the
cube. This suggests that the tails are scattered off the spacetime
curvature only once, picking up a single factor of mass.

In a second extension we considered non-spherically symmetric test
fields on the dynamical backgrounds generated by collapsing scalar
fields. We found that the exponent of the tails varies with multipole
index roughly as for fixed backgrounds. Furthermore we used the fact
that QN ringing is less damped in the higher modes to make precise
measurements of QN frequencies. For models in which the mass of the
background was reasonably constant we found excellent agreement with
the predictions on a fixed Schwarzschild background.  For models with
significant mass loss we found a shift in QN frequency corresponding
to the changing mass of the spacetime.

In summary, we found that the predictions for power-law tails of
perturbations of Schwarzschild spacetime \cite{Pr} hold to reasonable
approximation, even quantitatively, in a variety of situations to
which the predictions might seem initially not to apply.

In the interest of brevity and timeliness, the results reported here
do not exhaust the interesting questions that might be asked.  It will
be particularly interesting to explore in more detail the behavior of
power-law tails and QN ringing on the critical boundary between
collapsing and noncollapsing initial data. It should be said here that
our code is probably capable only of much coarser accuracy than that
of Choptuik \cite{Chop}, and will have to be modified for this
purpose. On the other hand, our code was adequate for verifying, for
our two families of initial data, one of Choptuik's crucial results:
that for marginal black hole formation the mass of the hole depends in
a universal way on the parameter of the family (in our case the
amplitude). See Fig.~15.

\thanks

We wish to thank Jeffrey Winicour and Roberto G\'omez for discussions
and for making available their results. This work was supported in
part by grant NSF PHY92-07225 and by research funds of the University
of Utah. J.P. acknowledges hospitality and support from the Institute for
Theoretical Physics at UCSB and the National Science Foundation grant
PHY89-04035.

\figure{Fig.~1: The conformal diagram of the spherical collapse
spacetime for a final hole mass $M_{f}$. Shown is our null grid in
relation to the future event horizon ${\cal H_+}$, future null
infinity $scrit_+$, future timelike infinity $i_+$, and spacelike infinity
$i_0$. Null grid lines pointing to the top left are lines of constant
$v$; those pointing to the top right are lines of constant $u$ or
$t_B$. Null data are set on the bottom right side of the grid at
$u=u_{0}$.   The curved
lines from left to right are a) $r=const.<2M_f$, b) $r=2M_f$ and c)
$r=const.>2M_f$.}

\figure{Fig.~2: Our null grid in coordinates $t(u,r)$ and $r$.
The $u=const.$ ($t_B=const.$) grid lines approach the event horizon in
a finite interval of $u$ which is an infinite interval of $t_B$.}

\figure{Fig.~3: The convergence of $\phi(r=10,t_B)$ with decreasing
grid size.  On the vertical axis is the log of the $l_2$-norm of the
error (compared to a very-small-grid numerical solution); on the
horizontal axis is the norm of the relative grid size. Data on
$u=u_{0}$ for this solution is a Gaussian $\phi(r)$ with center at
$1.0$, width $0.1$ and amplitude $0.06$.  The curves correspond to the
following sections: $a)0<t_B<10, \ b) 10<t_B<20,\ c) 20<t_B<30\
d)30<t_B<40\ e)40<t_B<50$. The slope of $a$ is $1.34\pm0.07$ and of
$d$ is $1.72\pm0.04$.}

\figure{Fig.~4: The convergence of $\phi(r,t_B=50)$ with decreasing
grid size.  The axes and the initial data are the same as for Fig.~3.
From top to bottom are $0<r<10$, $10<r<20$, $20<r<30$, $30<r<40$ and
$40<r<50$.  The slope of the top graph is $1.21\pm0.04$ and the slope
of the bottom graph is $1.70\pm0.04$. }

\figure{Fig.~5: Part of a plot of $\phi(r=10)$ versus $t_B$, in a
region where $\phi$ is dominated by QN ringing. From bottom
to top are shown  runs with initial radial grid sizes $1/20$,
$1/40$, $1/80$, $1/160$ and $1/320$. Initial data is the same as in
Fig.~3, which corresponds to the formation of a black hole with
negligible radiation of mass.}

\figure{Fig.~6: Results of the same runs  as in Fig.~5, but here in
a region where $\phi$ is dominated by the power-law tail.  Note that
the higher the precision, the smaller the value of $t_B$ where the run
stops due to an overflow, here at 53 and 57. From bottom to top the
curves correspond to initial radial grid sizes of $1/20$, $1/320$,
$1/160$, $1/80$, $1/40$.}

\figure{Fig.~7: Log-log plots of $\phi(r=10,t_B)$ for Gaussian data
with different amplitudes. Each plot starts after the last change of
sign of $\phi$. The amplitudes of the initial Gaussian are a) $0.06$,
b) $0.034$ (marginally collapsing), c) $0.033$ (marginally
noncollapsing), d) $0.01$.  The power law nature of all curves is
clearly visible; the exponents are a) $-2.74$, b) $-2.63$, c) $-2.63$,
d) $-2.68$, compared to a linearized theory prediction of $-3$.  Only
the two collapsing cases, show QN ringing.}

\figure{Fig.~8: A closeup on $\phi(r=10,t_B)$ in the region of QN
ringing for initial Gaussian data. Solid line: Amplitude $0.034$,
which collapses marginally, showing QN ringing (the small feature
around $t\approx5$).  Dotted line: Amplitude $0.033$, which marginally
does not collapse.  Note the complete absence of QN ringing.}

\figure{Fig.~9: A closeup of $\phi(r=10,t_B)$ in the region of
QN ringing, for static-static data a) Amplitude $0.50$, which
collapses, and here shows QN ringing. b) Amplitude $0.35$,
showing only faint QN ringing. c) Amplitude
$0.29$, which marginally collapses. d) Amplitude $0.28$ which
marginally does not collapse. There is no qualitative difference
between c) and d), in contrast to the case of Gaussian data depicted
in Fig.~8.} 

\figure{Fig.~10: Log-log plots of $\phi(r=10,t_B)$ for static-static
data with different initial amplitudes $\phi_{0}$. Each plot starts
after the last change of sign of $\phi$. The amplitudes are $a$) $0.5$,
$b$) $0.29$ (marginally collapsing), $c$) $0.28$ (marginally
noncollapsing), $d$) $0.01$.  The  power law nature of all curves is clearly
visible; the exponents are a) $-2.08$, b) $-1.98$, c) $-1.95$, d) $-1.86$,
compared to a linearized theory prediction of $-2$.
The power laws are a) $-2.08$, b) $-1.98$, c) $-1.95$, d) $-1.86$,
compared to a linearized theory prediction of $-2$.
Only curve $a$ shows clear QN
ringing.}

\figure{Fig.~11:} Test fields $\varphi_l^m(r=10,t_B)$ on a Gaussian
background, amplitude $0.06$. The three graphs are $l=0$, $l=1$, $l=2$
(in order of increasing frequency).

\figure{Fig.~12: Log-log plots of test fields $\varphi_l^m(r=10,t_B)$ for
different multipole indices $l$ on one background spacetime. The
background is evolved from Gaussian initial data for $\phi$ with
amplitude $0.02$ (noncollapsing). The initial data for the test field
is in each case a Gaussian (the test field amplitude, $1.0$, has no
significance). From top to bottom the curves correspond to $l=0, 1, 2,
3$.  The best fit for the power law exponents are $-2.77$, $-3.95$,
$-5.94$, $-8.34$, compared to predictions of $-3$, $-5$, $-7$ and $-9$.
}

\figure{Fig.~13: Log-log plots of test fields $\varphi_l^m(r=10,t_B)$
for different multipole indices $l$ on one background spacetime. The
background is evolved from static-static data for $\phi$ with
amplitude $0.1$ (noncollapsing). The initial data for the test field
is in each case a Gaussian (the test field amplitude, $1.0$, has no
significance). From top to bottom $l=0$, then $l=1$, $l=2$ and $l=3$.
The best fit for the power laws exponents are $-2.70$, $-3.66$,
$-5.52$, $-7.26$, compared to predictions of $-3$, $-5$, $-7$ and
$-9$.}

\figure{Fig.~14: Log-log plots of the initial Bondi mass and the
amplitude of the $t^{-3}$ power-law tail $\phi(r=10,t_B)$,
both versus the amplitude of
initial data. The upper graph of each pair (full dots) represents
Gaussian data, the other (empty dots) static-static data
The region of dense data points marks the
collapse-noncollapse transition on each graph.
Both the initial Bondi mass and the amplitude of the tail scale 
rather precisely as powers of the initial data amplitude for
noncollapsing data. The powers are, for the mass 1.99 (Gaussian) and 
1.97 (static-static), and for the tail 3.22 (Gaussian) and 
3.01 (static-static).}

\figure{Fig. 15: Mass of black hole formed versus difference between
the amplitude of the initial data and the critical amplitude. On the
vertical axis, $\ln({\rm mass})$. On the horizontal axis
$\ln[(p-p_*)/p_*]$, where $p$ is $\phi_0$ or $A$ and $p_*$ its
critical value. Empty squares denote static-static data, full squares
denote Gaussian data. The fact that the static-static data points are
further to the left means that they were measured closer to
criticality. This was necessary because the universal power-law
behaviour of the mass develops only closer to criticality for
static-static than for Gaussian data. The deviation from this
behaviour is clear in the plot for the empty squares for
$(\phi_0-\phi_{0*})/\phi_{0*}>e^{-6}$. For smaller values the empty
squares fall approximately on a straight line.  The sets were shifted
vertically (but not horizontally as was done in Ref.  \cite{Chop})
with respect to each other in order to place them on one line. This
corresponds to an overall constant multiplicative factor in the black
hole mass which is not universal but depends upon the family of
initial data (Gaussian or static-static). We assumed $A_*=0.03280$ and
$\phi_{0*}=0.2860$.  Formally, the best fit to the slope (power-law) for
Gaussian data alone is $0.39$, for static-static data alone $0.31$,
and combining both sets $0.35$. A full analysis of the uncertainties
in these slopes has not been carried out, but we assume that within
numerical accuracy the slopes agree with each other and with the slope
$0.37$ found by Choptuik \cite{Chop}.}

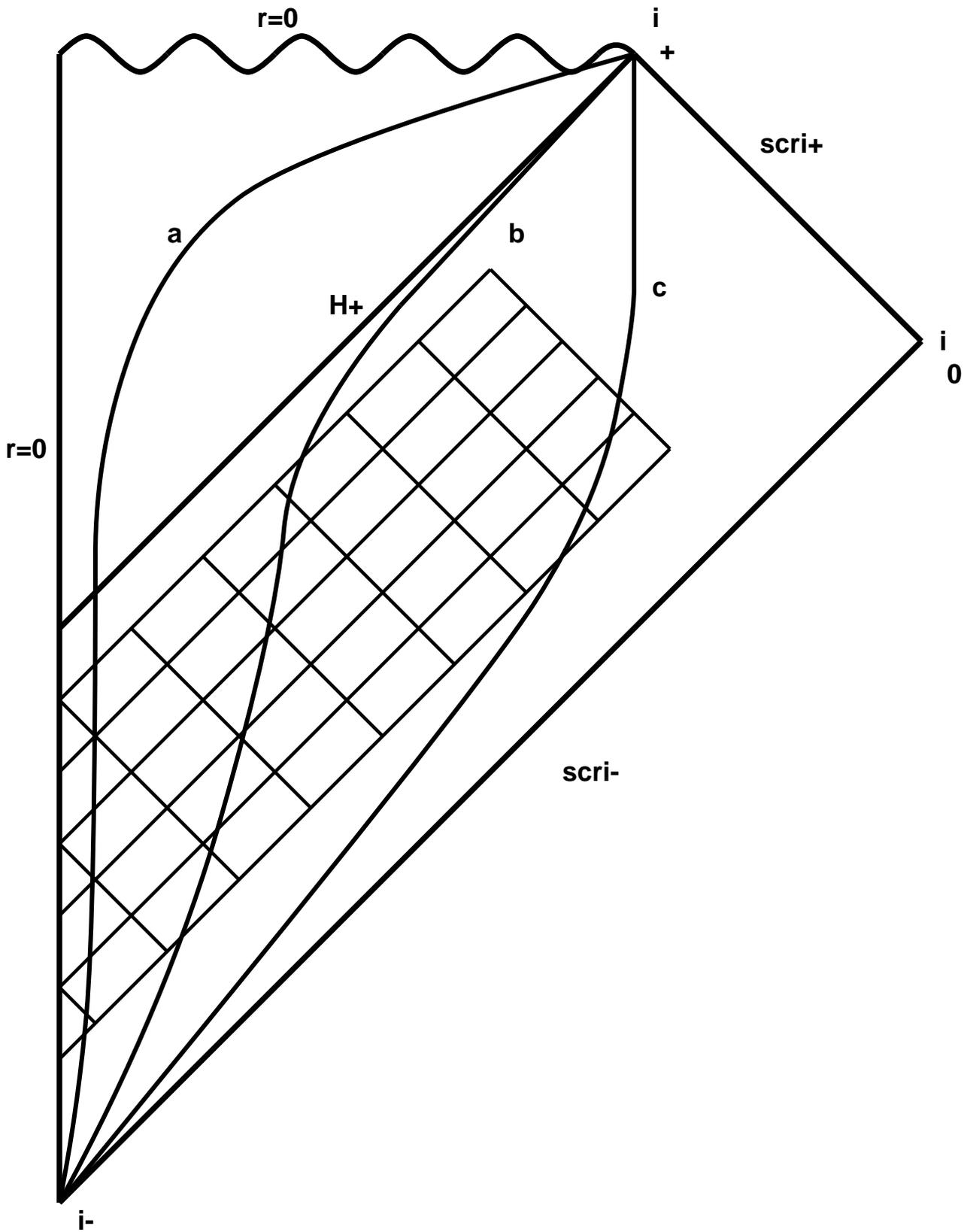
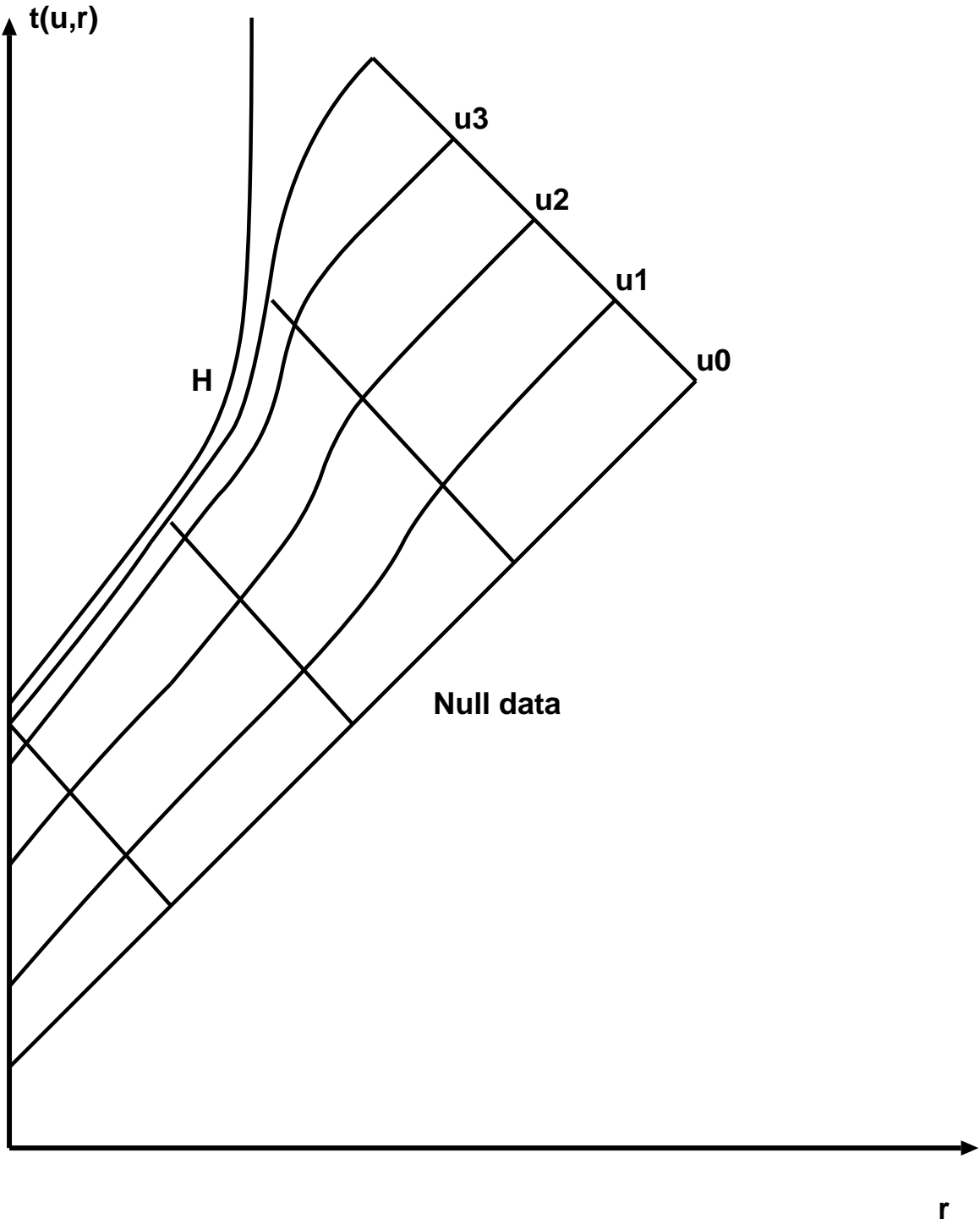
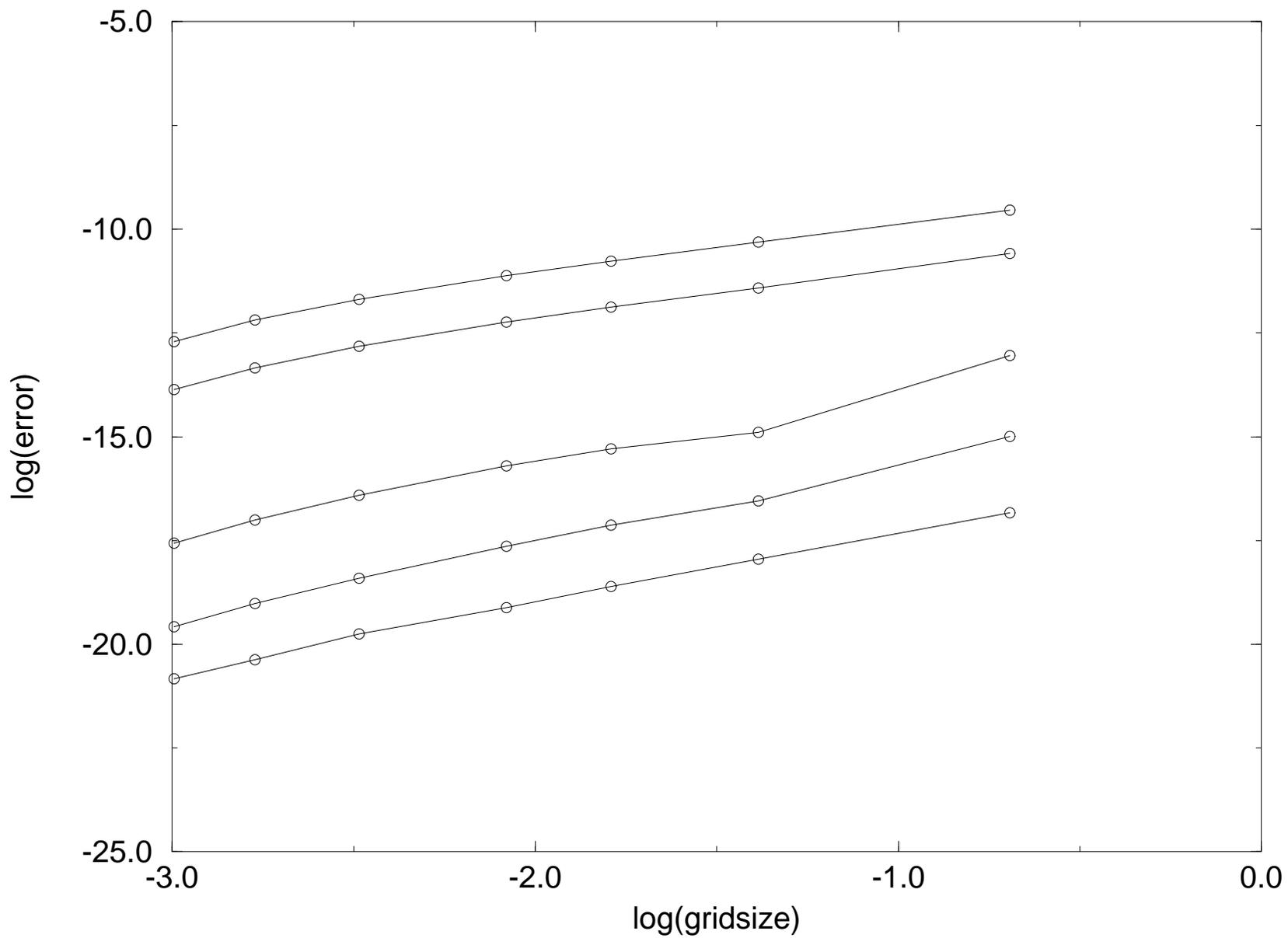
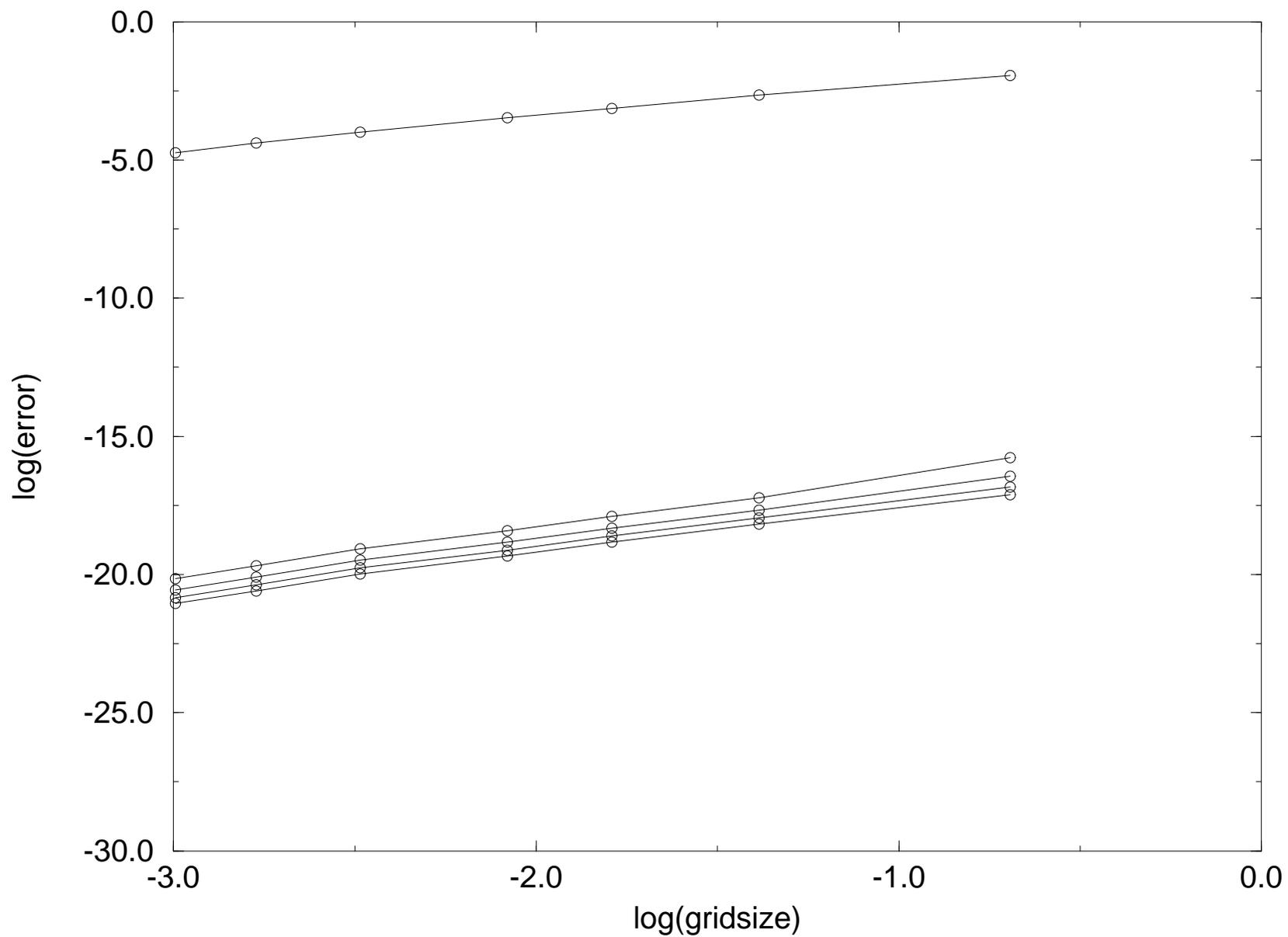
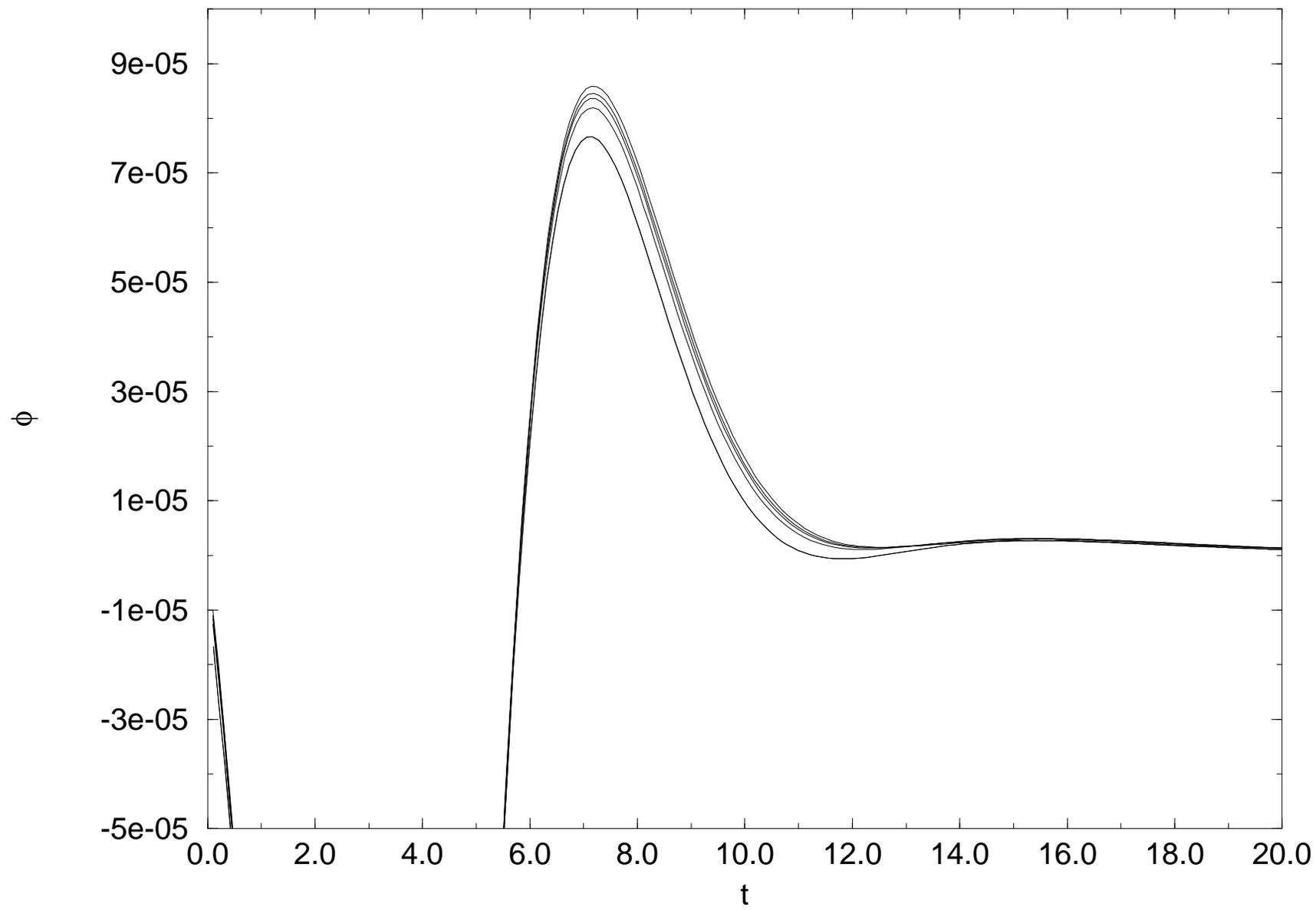
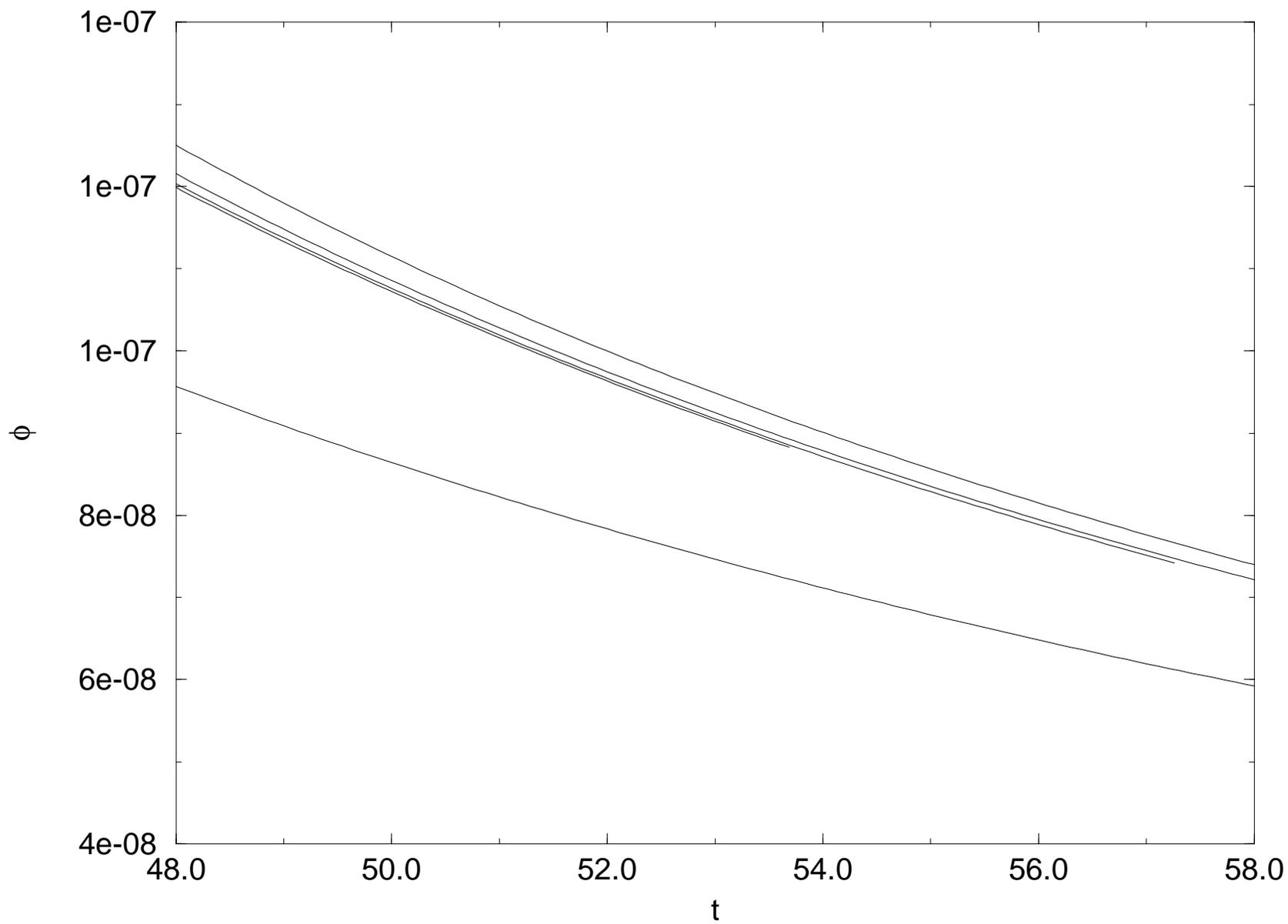
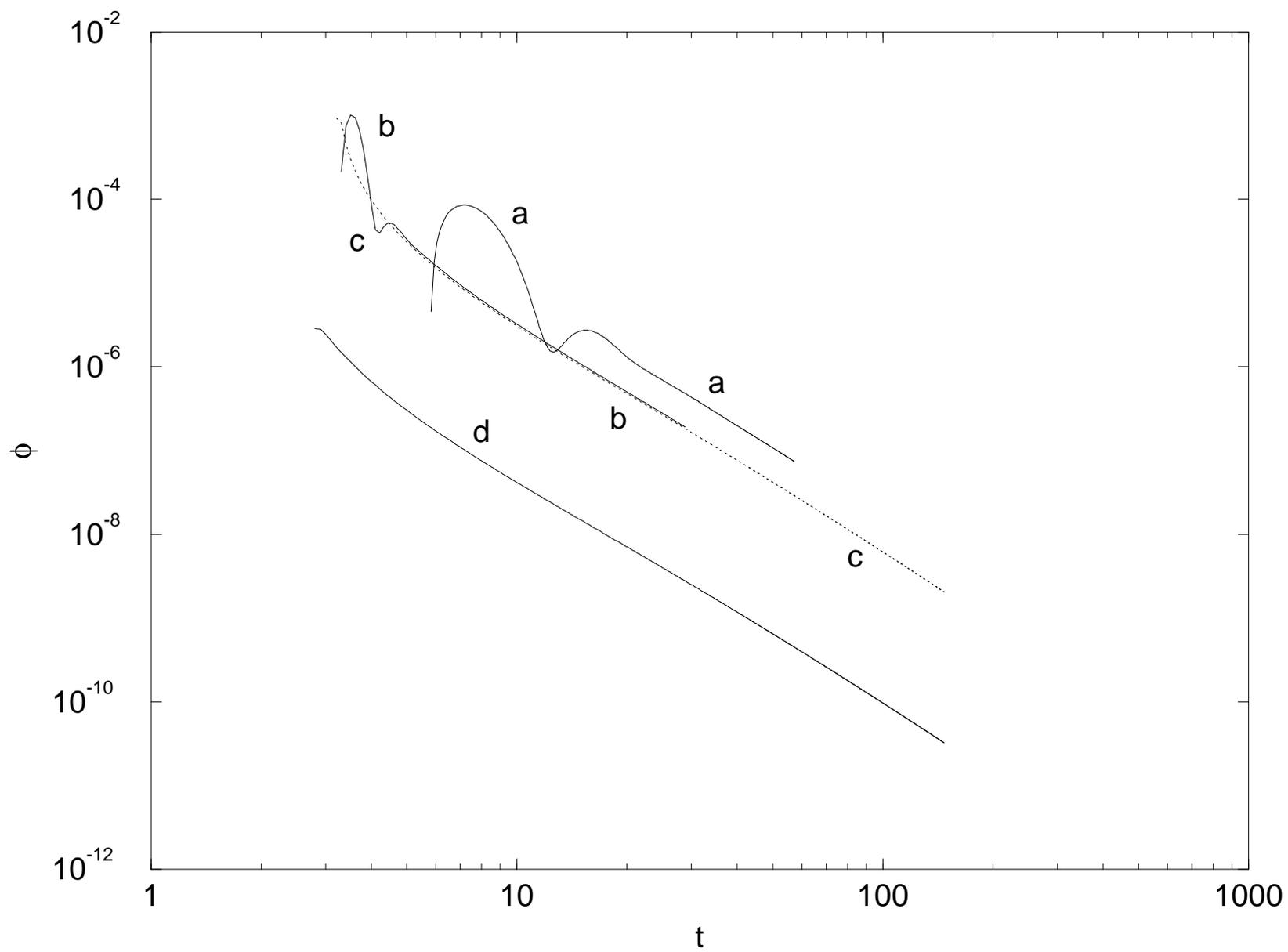
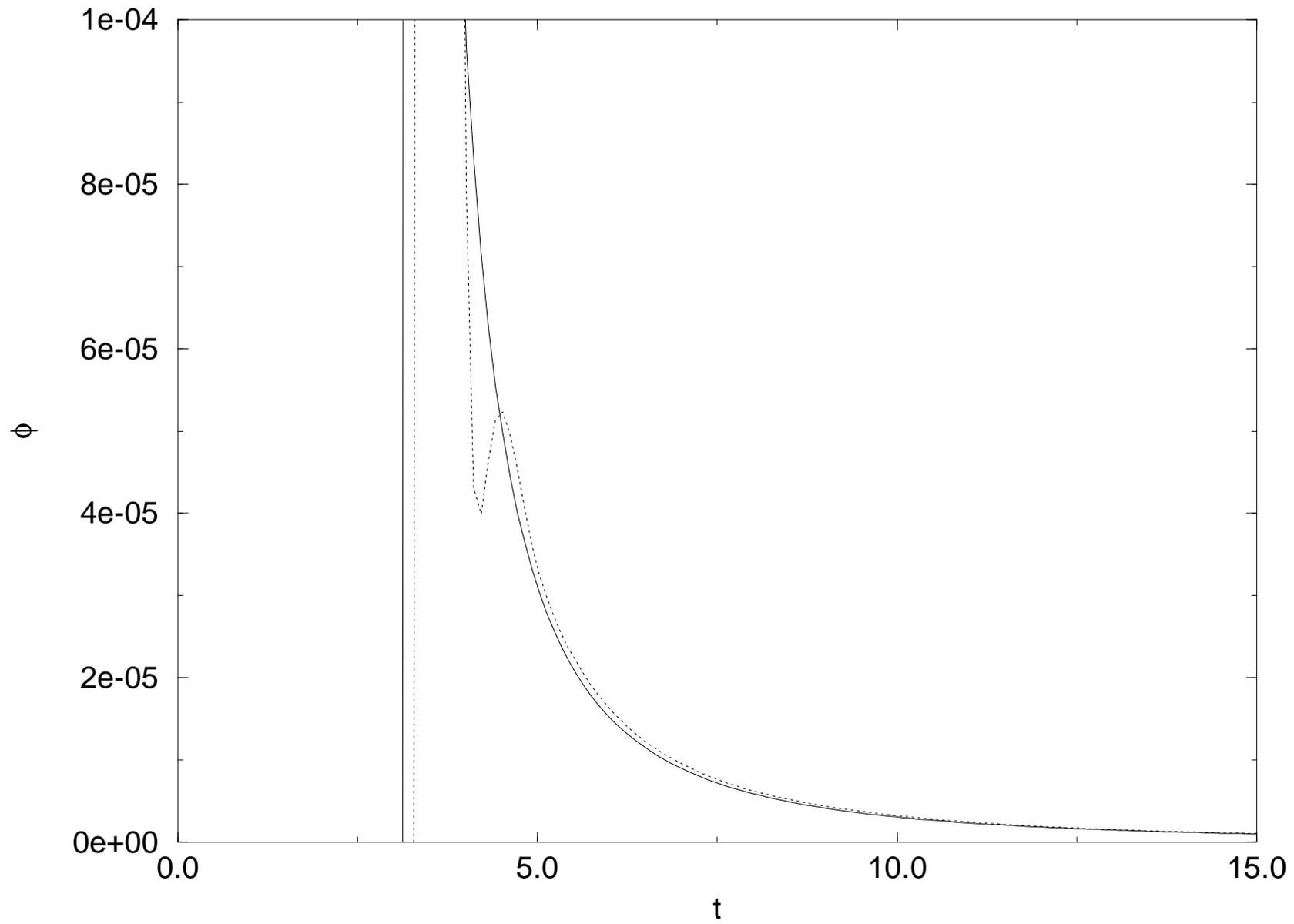
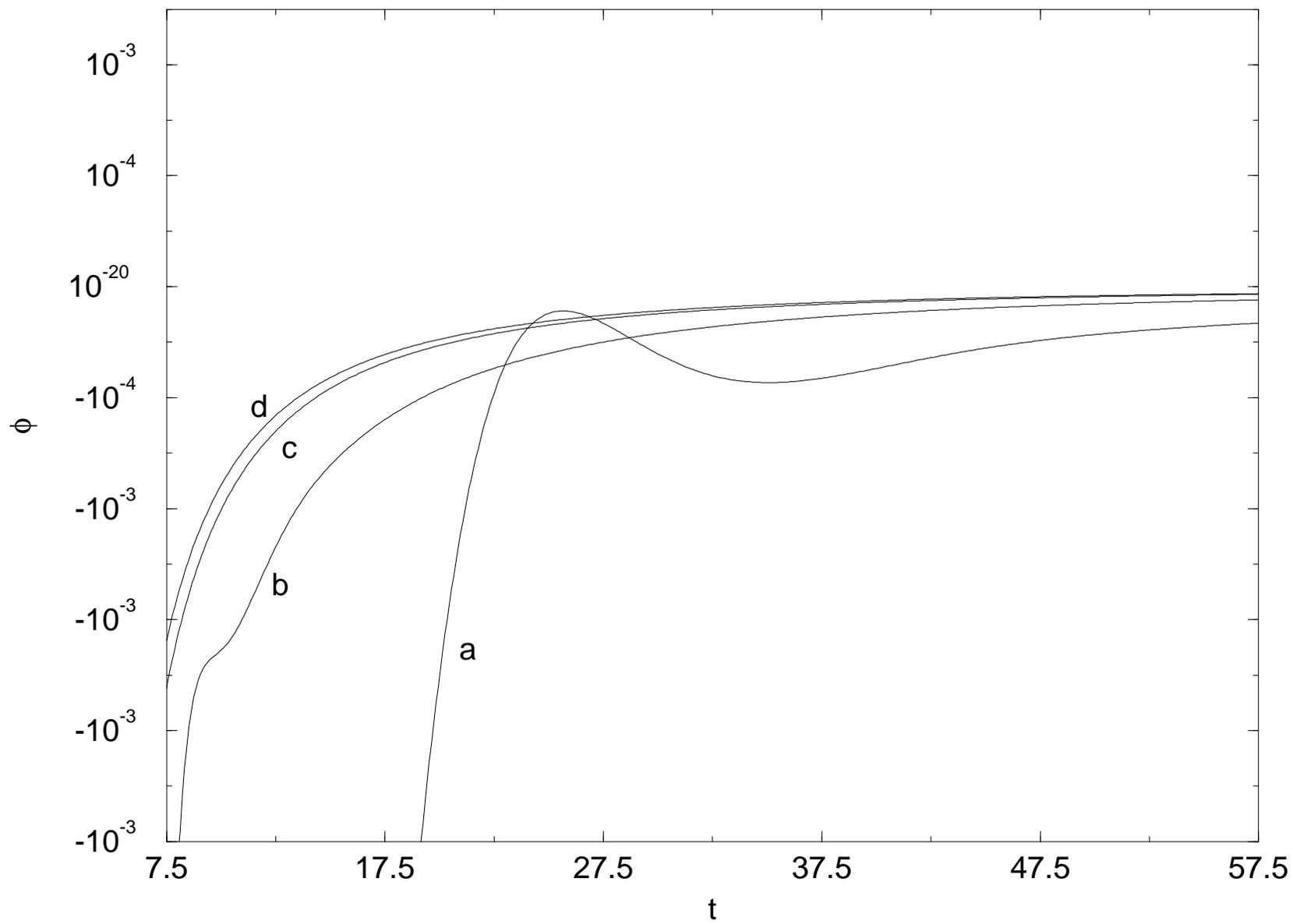
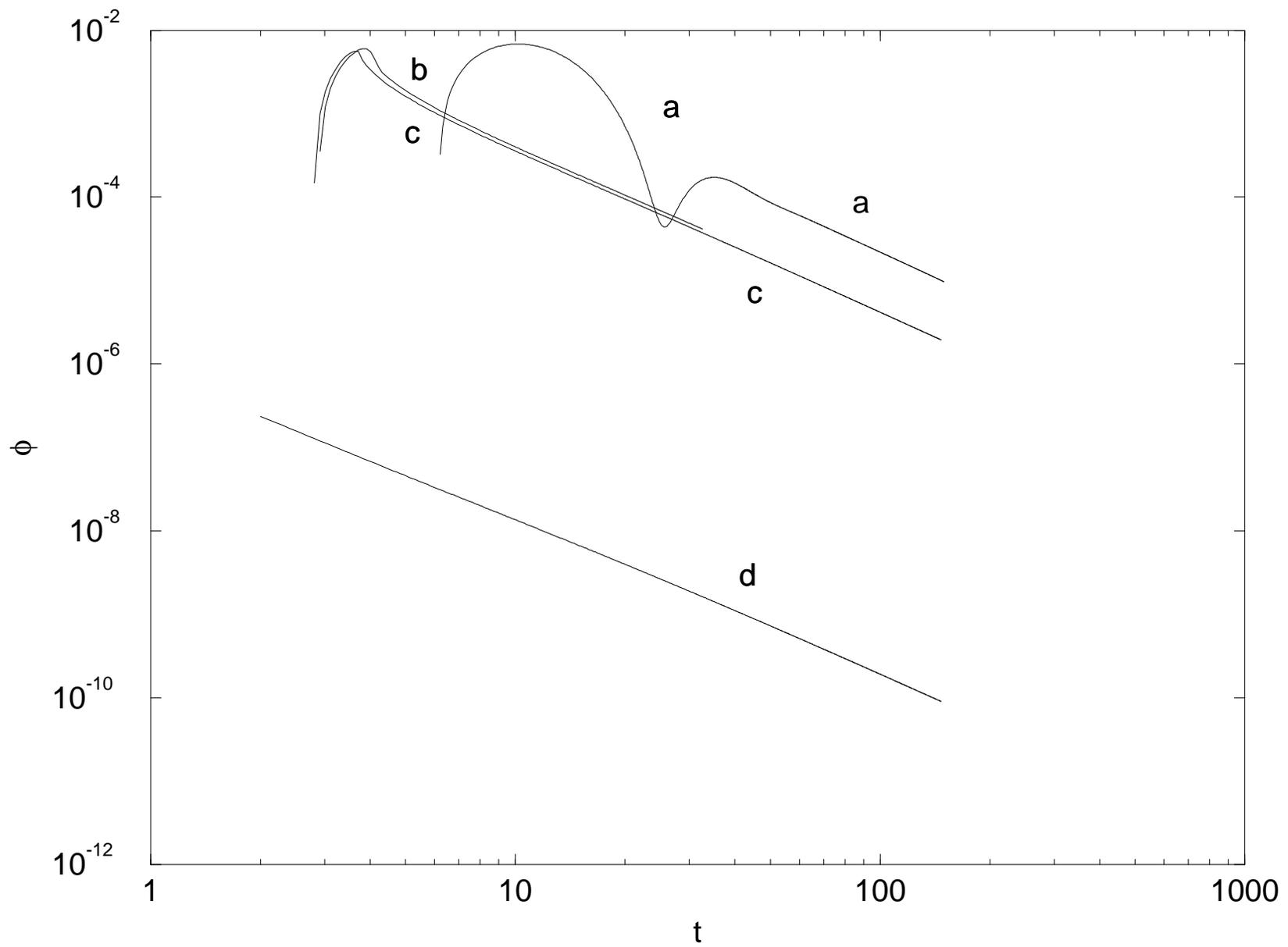
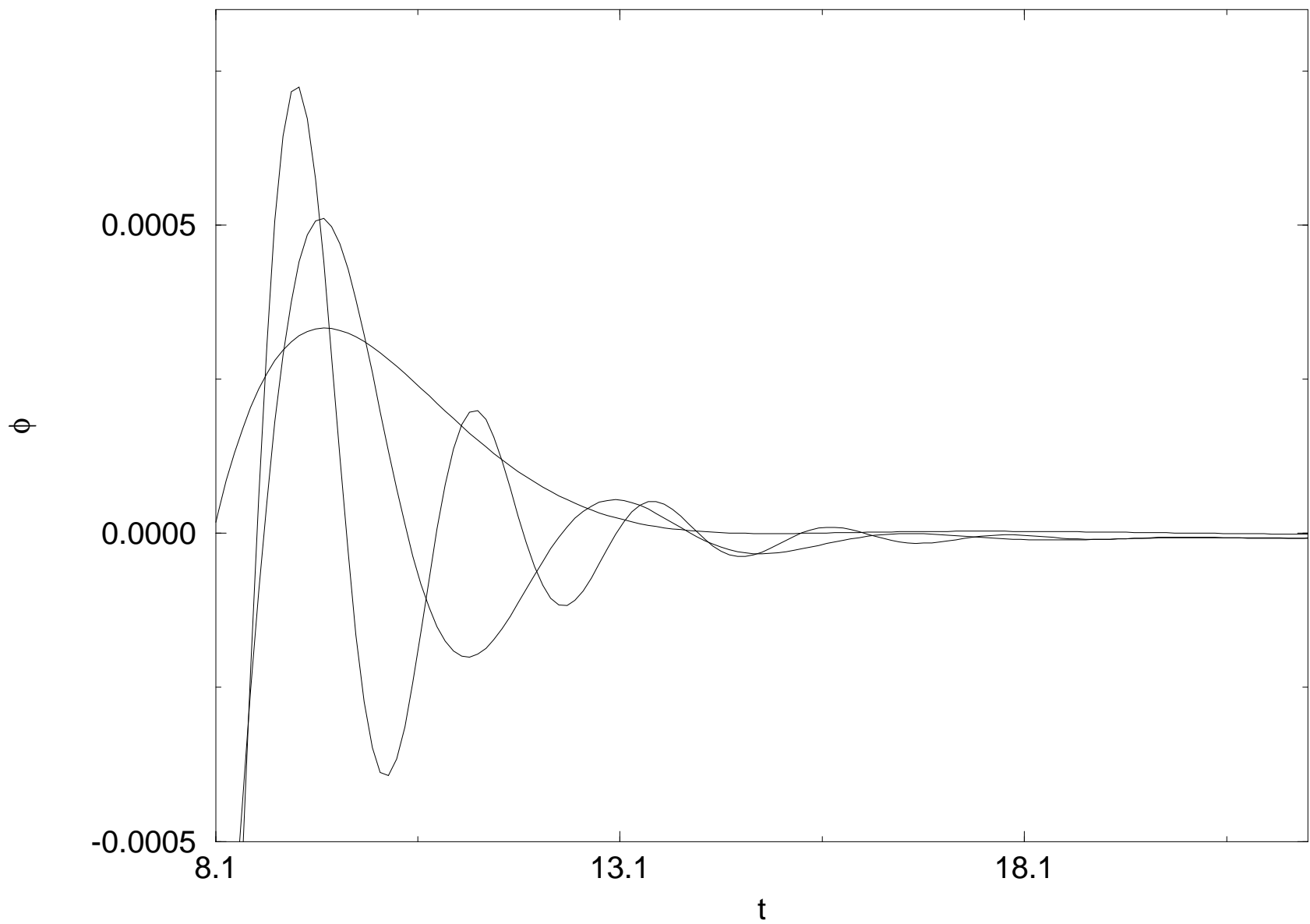
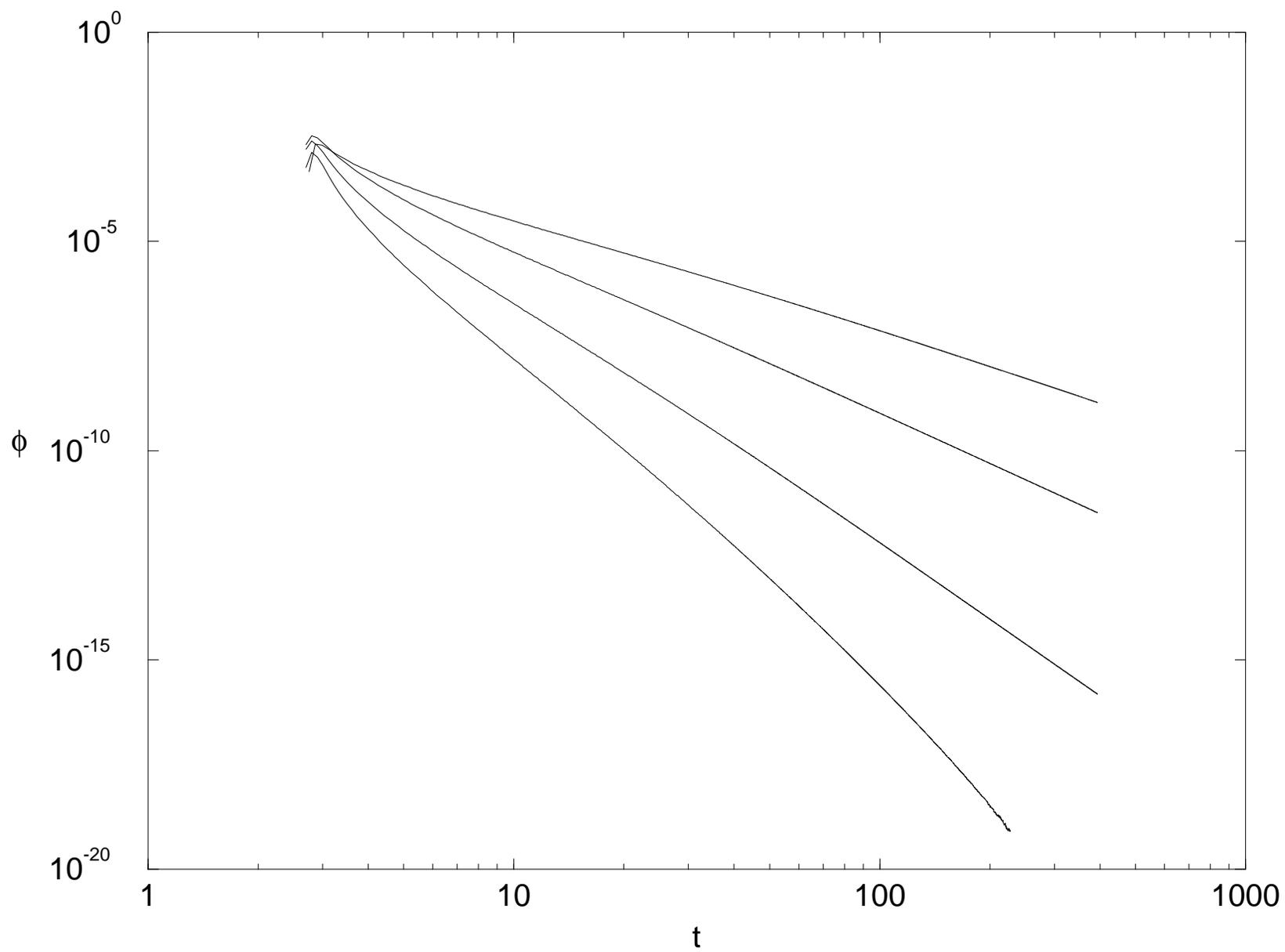
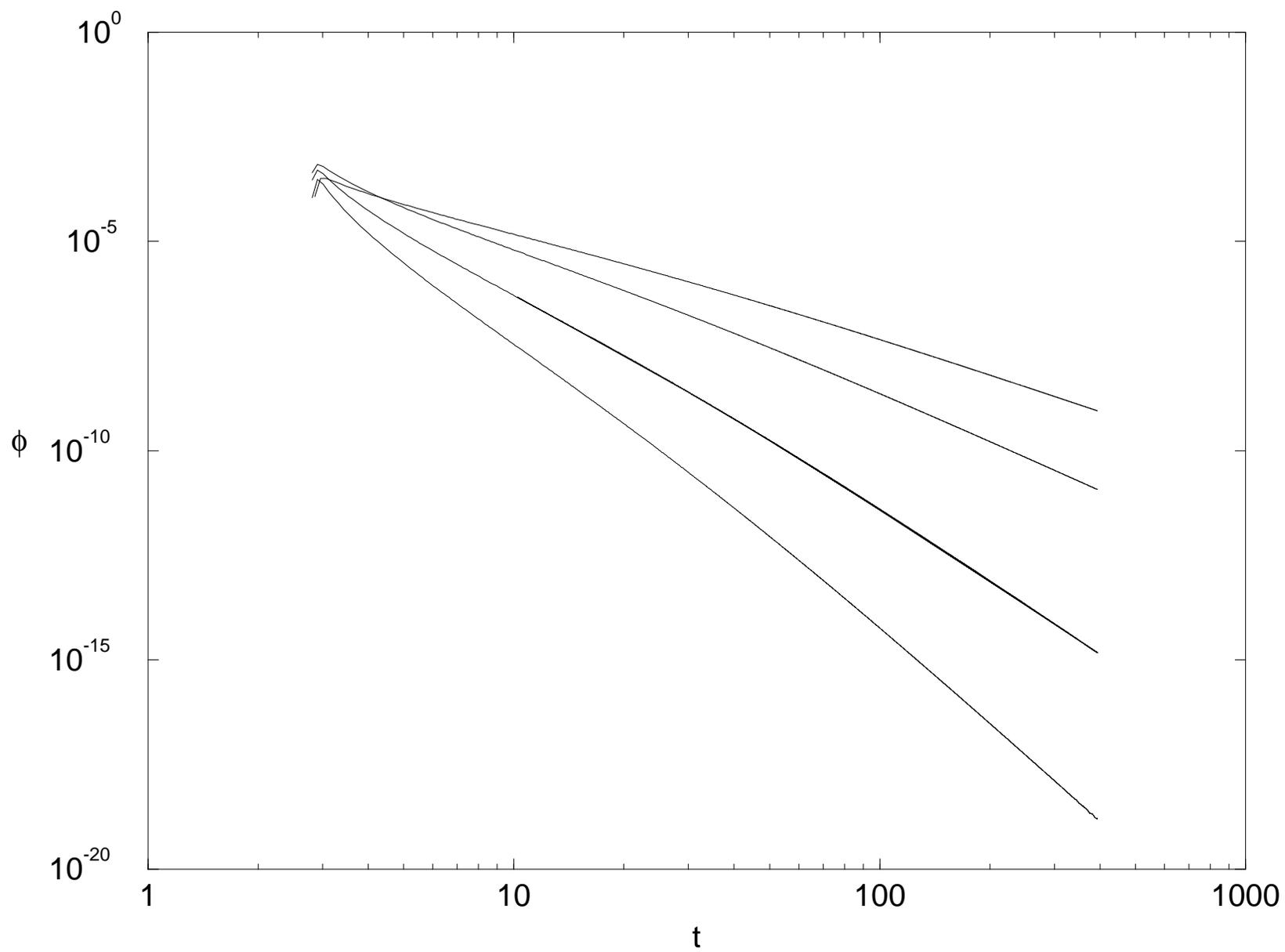
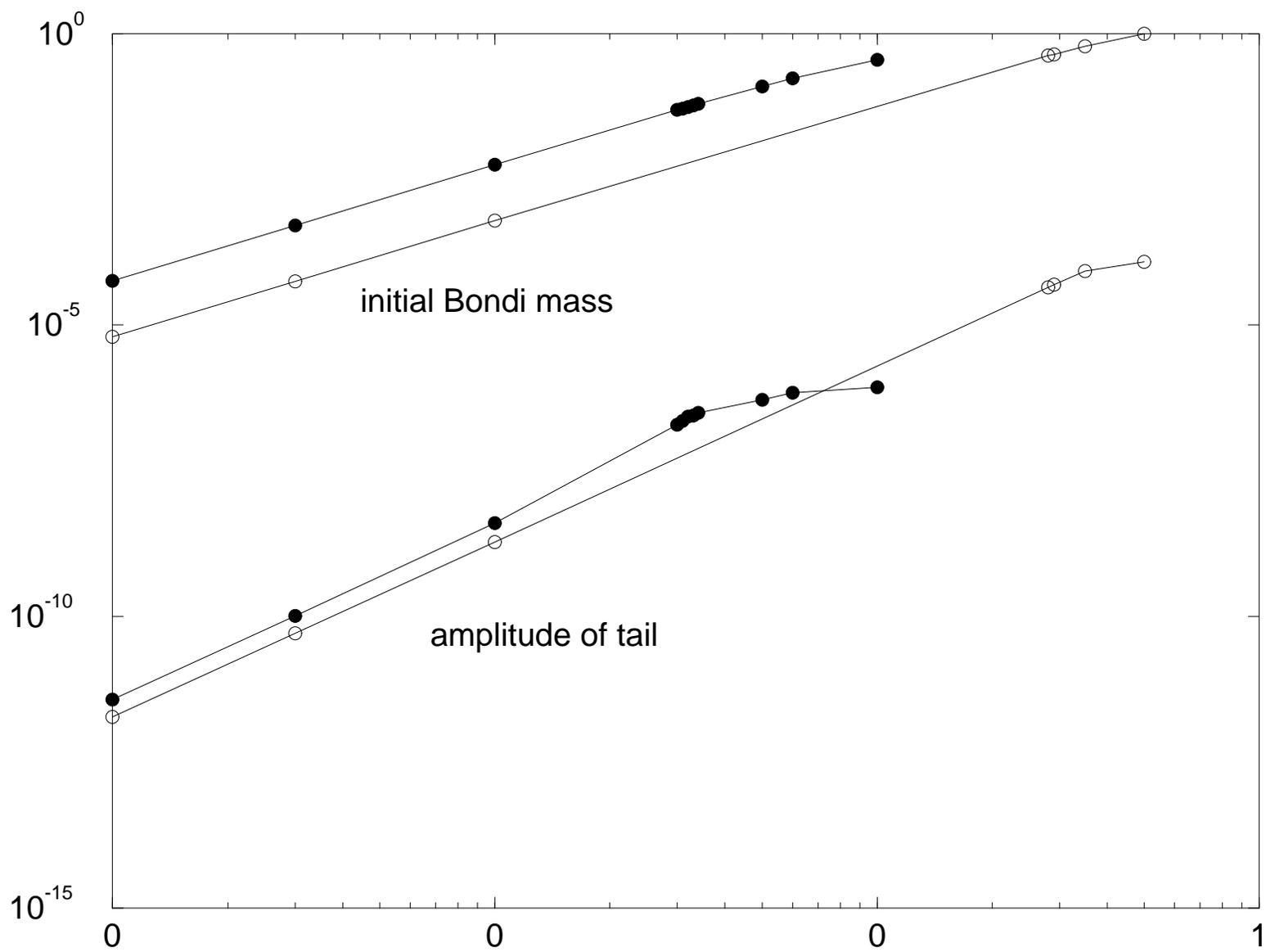
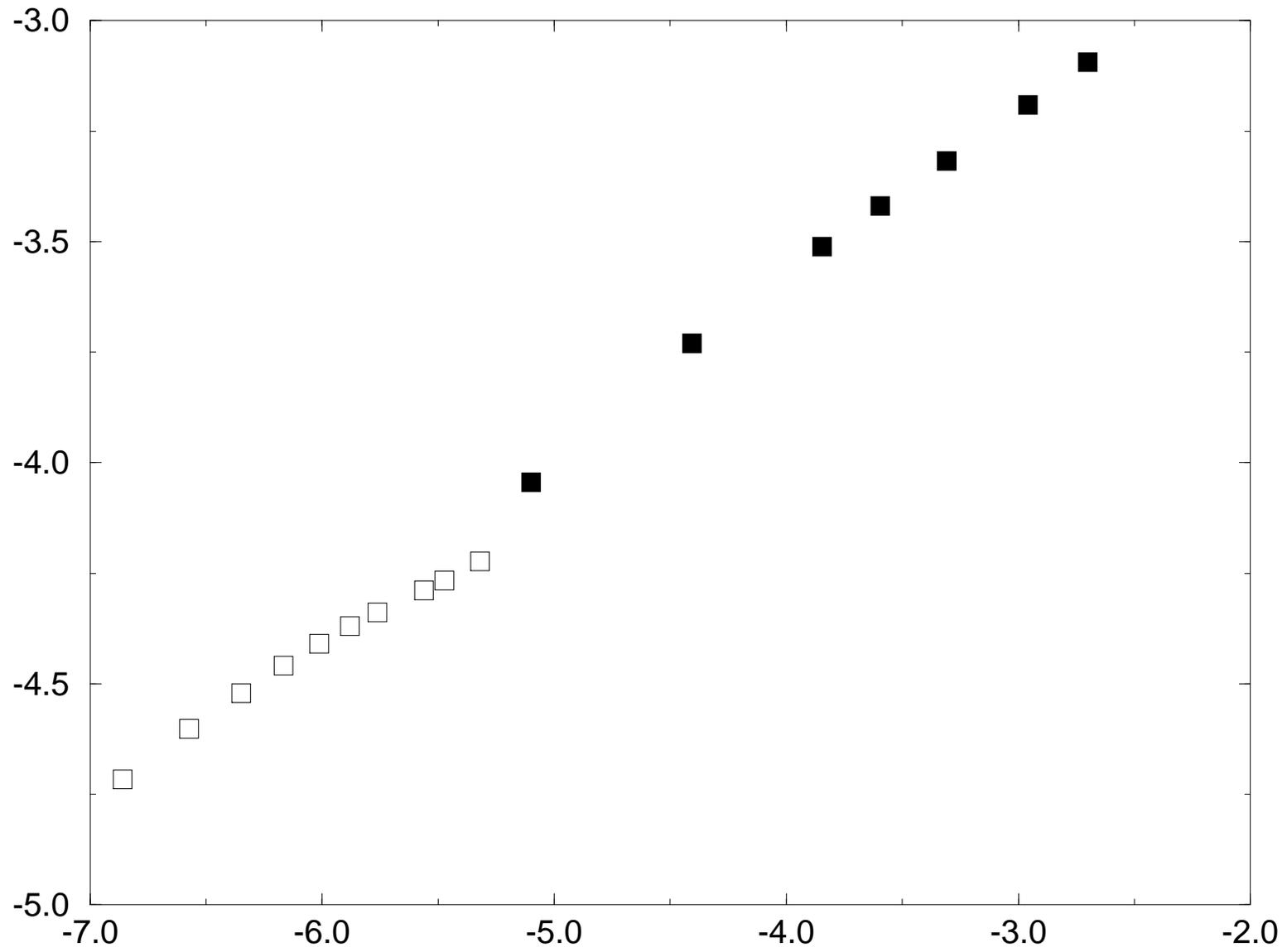
\end{document}